\documentclass[aps,prd,twocolumn,a4paper]{revtex4}
\usepackage{color}
\usepackage{graphicx}
\usepackage{epsfig}

\usepackage{longtable,lscape}
\usepackage{amssymb}

\def\nn {\nonumber}
\newcommand{\be}{\begin{equation}}
\newcommand{\ee}{\end{equation}}
\newcommand{\bea}{\begin{eqnarray}}
\newcommand{\eea}{\end{eqnarray}}

\newcommand{\ps}{p \!\!\! /}

\newcommand{\del}{\partial}

\begin{document}
\title{Diffusion of $\Lambda_c$ in hot hadronic medium and its impact on $\Lambda_c/D$ ratio} 
\author{Sabyasachi Ghosh$^{1}$, Santosh K Das$^{2,3}$, Vincenzo Greco$^{2,3}$, 
Sourav Sarkar$^{4}$, Jan-e Alam$^{4}$}

\affiliation{$^1$Instituto de Física Teórica, 
Universidade Estadual Paulista, Rua Dr. Bento Teobaldo Ferraz, 271 — Bloco II, 
01140-070 Sao Paulo, SP, Brazil}

\affiliation{$^2$ Department of Physics and Astronomy, University of Catania, 
Via S. Sofia 64, 1-95125 Catania, Italy}
\affiliation{$^3$ Laboratori Nazionali del Sud, INFN-LNS, Via S. Sofia 62, I-95123 Catania, Italy}

\affiliation{$^4$Theoretical Physics Division, 
Variable Energy Cyclotron Centre, 1/AF, Bidhan Nagar, 
Kolkata - 700064}

\begin{abstract}
The drag and diffusion coefficients of the $\Lambda_c(2286$ MeV) have been evaluated in the 
hadronic medium which is expected to be formed in  the later stages of the evolving fire ball produced in
heavy ion collisions at RHIC and LHC energies. 
The interactions between the $\Lambda_c$ and the hadrons in the medium 
have been derived from an effective hadronic Lagrangian as well as from the 
scattering lengths, obtained in the framework of heavy baryon 
chiral perturbation theory (HB$\chi$PT). 
In both the approaches, the magnitude of the transport coefficients
are turn out to be  significant.  A larger value is obtained in the former approach
with respect to the latter.
Significant values of the coefficients indicate substantial 
amount of interaction of the $\Lambda_c$ with the hadronic thermal bath. Furthermore, the transport coefficients 
of the $\Lambda_c$ is found to be different from the transport coefficients of $D$ meson. 
Present study 
indicates that the hadronic medium has a significant impact on the $\Lambda_c/D$ ratio in 
heavy ion collisions. 

\vspace{2mm}

\end{abstract}
\maketitle

\section{Introduction}
One of the  primary aims of the ongoing nuclear collision programmes at 
the Relativistic Heavy Ion Collider (RHIC) and Large Hadron Collider (LHC) energies 
is to create a new state of matter known as Quark Gluon Plasma (QGP),  
the bulk properties of which  are governed by the light 
quarks and gluons.  
Heavy quarks (HQs\,$\equiv$  charm and beauty) play crucial roles to
understand the properties of QGP~\cite{hfr}, because they  can  witness the 
entire space-time evolution of the system as they are produced in the initial hard collision 
and remain extant during the evolution. Heavy flavor as a probe of the medium has generated
significant interest in the recent past due to the suppression of its momentum distribution
at large momentum in the thermal medium, denoted by 
$R_{AA}(p_T)$ ~\cite{stare,phenixelat,phenixe,alice} and 
its elliptic flow ($v_2$)~\cite{phenixe}. Several attempts have been made 
to study these factors  within the framework of 
Fokker Plank equation~\cite{hfr,moore,rappv2,hvh, 
hiranov2,cmko,Das,alberico,jeon,bass,rappprl,ali,hees,qun} 
and Boltzmann equation~\cite{gossiauxv2,gre,you,fs}. 
However, the roles of hadronic phase have been ignored in these works.

In heavy ion collision (HIC) at ultra-relativistic energies 
the appearance of the hadronic phase is inevitable.  
To make reliable characterization of the QGP  
the role of the hadronic phase should be assessed 
and its contribution must be subtracted out from the data.
Recently the diffusion coefficient of the 
$D$ and $B$ mesons have been evaluated in the hadronic phase~\cite{laine,MinHe,Ghosh,abreu,Das3,abreu1,juan,juan1} 
and their effects 
on $R_{AA}(p_T)$ at large transverse momentum ($p_T$)
~\cite{prdh} and elliptic flow ($v_2$)~\cite{tx,frr} has been studied and found to be significant. 
Apart from the heavy mesons ($D$ and $B$) the heavy baryon ($\Lambda_c$) is also 
significant as its enhancement~\cite{lee,lee1} due to quark coalescence would 
affect the $R_{AA}(p_T)$ of non photonic electrons. Furthermore the 
baryon-to-meson ratio, ($\Lambda_c/D$), is fundamental for the understanding of in medium 
hadronisation~\cite{vg} with respect to the light quark sector~\cite{vg1}.
Enhancement of heavy baryon-to-meson 
ratio ($\Lambda_c/D$) in Au+Au collisions compared to p+p collisions affects  
the non-photonic electron spectrum ($R_{AA}$)~\cite{soren,fra,max,greco}. 
The branching ratio for the process $\Lambda_c \rightarrow e + X (4.5\% \pm 1.7\%)$ is 
smaller than  $D \rightarrow e + X (17.2\% \pm 1.9\%)$, resulting in less electrons 
from decays of $\Lambda_c$ than D. Hence, enhancement of $\Lambda_c/D$ ratio 
in Au+Au collision will reduce the observed non-photonic electrons. 
We notice that the $p_T$ dependence of the 
$\Lambda_c/D$ ratio may get modified 
further in the hadronic medium as their  interactions with hadrons are 
non-negligible.  Keeping this in mind we attempt to study 
the transport coefficients (drag and diffusion coefficients) of $\Lambda_c$ in 
hadronic phase. 

The paper is organized as follows. In the next section we discuss the formalism
used to evaluate the drag and diffusion coefficients of the heavy flavored baryon in a hot hadronic
matter. Section III is devoted to present the results.
Section IV contains summary and discussions.

\section{Formalism}
The drag and diffusion coefficients of the charmed baryon $\Lambda_c$,
propagating through a hot hadronic medium have been evaluated
using the scattering length obtained in Ref.~\cite{Liu_Zhu},
where Liu et. al have estimated Next-to-Next-to-Leading order (NNLO) amplitudes 
in the framework of Heavy Baryon Chiral Perturbation Theory $(HB\chi PT)$.
We consider the elastic interaction of $\Lambda_c$ with 
thermal pions, kaons and $\eta$ mesons.  
The temperature of the bath can vary from $T_c(\sim 170)$ to $T_F(\sim 120$ MeV),
relevant for heavy ion collisions at RHIC and LHC energies. Here $T_c$ is the transition temperature 
at which the QGP formed in HIC makes a transition to hadrons and 
$T_F$ is the freeze-out temperature at which the hadrons cease to interact. 
In this temperature 
range the abundance of $\Lambda_c$ and $D$  is  
small and  their thermal production and annihilation can be ignored.
Hence, in evaluating the  drag and diffusion coefficients of the $\Lambda_c$ 
only elastic processes will be considered.

For the elastic scattering  of $\Lambda_c$ of momentum $p_1$ with
a thermal hadron, $H$ of momentum $p_2$ {\it i.e} for the process, 
$\Lambda_c(p_1) + H(p_2) \rightarrow \Lambda_c(p_3) + H(p_4)$, 
the drag coefficient, $\gamma$ can be expressed as~\cite{BS}(see also ~\cite{DKS,vc} ) :
\begin{equation}
\gamma=p_iA_i/p^2
\end{equation}
where $A_i$ takes the form 
\bea
A_i&=&\frac{1}{2E_{p_1}} \int \frac{d^3p_2}{(2\pi)^3E_{p_2}} \int \frac{d^3p_3}{(2\pi)^3E_{p_3}}
\int \frac{d^3p_4}{(2\pi)^3E_{p_4}}  
\nonumber \\ 
&&\frac{1}{g_{\Lambda_c}} 
\sum  {\overline {|{\cal{M}}|^2}} (2\pi)^4 \delta^4(p_1+p_2-p_3-p_4) 
\nonumber \\
&&{f}(p_2)\{1\pm f(p_4)\}[(p_1-p_3)_i] \equiv \langle \langle
(p_1-p_3)\rangle \rangle
\nn\\
\label{eq1}
\eea
where $g_{\Lambda_c}$ denotes the statistical degeneracy of $\Lambda_c$,
$f(p_2)$ is the Bose-Einstein (BE) or Fermi-Dirac (FD) distribution function 
depending upon the spin of $H$
in the initial channel. Similarly, the factor  $1 \pm f(p_4)$ represents
Bose enhanced or Pauli suppressed probability of the corresponding 
$H$ in the final channel.
${\overline {|{\cal{M}}|^2}}$ represents the modulus square of the spin averaged matrix
element for $\Lambda_c+H$ elastic scattering process. 
Eq.(\ref{eq1}) illustrate that the drag coefficient is the 
measure of the thermal average of the momentum transfer, $p_1-p_3$  weighted  by 
the interaction through 
$\overline{|{\cal{M}}|^2}$.

Similarly the diffusion coefficient $B$ can be expressed as:
\begin{equation}
B=\frac{1}{4}\left[\langle \langle p_3^2 \rangle \rangle -
\frac{\langle \langle (p_1\cdot p_3)^2 \rangle \rangle }{p_1^2}\right]
\label{eq3}
\end{equation}

With an appropriate choice of $T(p_3)$ both the drag  and diffusion coefficients 
can be expressed in a single equation as follows: 
\bea
&&\ll T(p_1)\gg=\frac{1}{512\pi^4} \frac{1}{E_{p_1}} \int_{0}^{\infty} 
\frac{p_2^2 dp_2 d(cos\chi)}{E_{p_2}} 
\nn\\
&&~~~\hat{f}(p_2)\{ 1\pm f(p_4)\}\frac{\lambda^{\frac{1}{2}}(s,m_{p_1}^2,m_{p_2}^2)}{\sqrt{s}} 
\int_{1}^{-1} d(cos\theta_{c.m.})  
\nonumber \\ 
&&~~~~~~~~~~~~~~~~~~
\frac{1}{g} \sum  {\overline {|{\cal{M}}|^2}} \int_{0}^{2\pi} d\phi_{c.m.} T(p_3)
\label{transport}
\eea
where $\lambda(s,m_{p_1}^2,m_{p_2}^2)=\{s-(m_{p_1}+m_{p_2})^2\}\{s-(m_{p_1}-m_{p_2})^2\}$ 
is the triangular function.

The drag and diffusion coefficients of $\Lambda_c$ can  be evaluated
in hadronic matter by using  ${\overline {|T|^2}}$ ~\cite{Das3}
in place of ${\overline {|{\cal{M}}|^2}}$ in Eq.~(\ref{transport}),
where the momentum independent $T$-matrix elements simply estimate
the strength of the $\Lambda_c$ interactions with the thermal hadrons. 

The scattering lengths of $\Lambda_c$ with light pseudoscalar
mesons $M=\pi, K, \overline{K}$ and $\eta$ 
have recently been obtained by Liu et. al \cite{Liu_Zhu} in the framework of
HB$\chi$PT.
From the scattering lengths, $a$ (say) of $\Lambda_c$ interacting with $M$, 
we can extract the $T$-matrix element by using the relation
\be
T=4\pi(m_{\Lambda_c}+m_M)a~.
\ee
where $m_{\Lambda_c}$ and $m_M$ are the masses of $\Lambda_c$ and mesons ($M$) respectively.  
From the scattering lengths ($a$ in fm), the extracted values of $T$  are given 
in Table I.
\begin{table}[h]
\begin{center}
\begin{tabular}{|c|c|c|}
\hline
& &  \\
$\Lambda_cM$ & $a ({\rm fm})$ & T   \\
& &  \\
\hline
& & \\
$\Lambda_c\pi$ & 0.06 & 9.28 \\
& & \\
$\Lambda_cK$ & -0.032$\pm$0.038 & -12.42$~{\rm to}~$1.06 \\
& & \\
$\Lambda_c\overline{K}$ & (0.79+0.27i)$\pm$0.044 & (72.75+47.9i)$~{\rm to}~$(207.6+47.9i)  \\
& &  \\
$\Lambda_c\eta$ & (0.35+0.19i)$\pm$0.044 & (55.27+34.32i)$~{\rm to}~$(71.17+34.32i)  \\
& & \\
\hline
\end{tabular}
\caption{Table showing the extracted values of T-matrix from the
scattering length, $a$, which are obtained by 
Liu et. al \cite{Liu_Zhu}.}
\label{table}
\end{center}
\end{table}

In Ref.~\cite{Liu_Zhu} Liu et al., have fixed
the LECs (low energy constants) with the help of relations based on  
quark model symmetry, heavy quark spin symmetry, $SU(4)$ flavor
symmetry as well as some empirical relations. However, a dimensionless
constant $\alpha'$ remains unknown, which is taken in the natural range
$[-1,1]$~\cite{Liu_Zhu,Guo}. Therefore, the table shows a band of numerical 
values of $T$-matrices, which are corrected up to the 
3rd order (${\cal O}(\epsilon^3)$) with the explicit power counting in HB$\chi$PT.
Using the $T$-matrices from Table~\ref{table} and corresponding BE
distributions for $H=M=\pi,~K,~\eta$ in Eq.~(\ref{transport}), we can get
an estimate of drag and diffusion coefficients of $\Lambda_c$ in hadronic
matter.

Besides the scattering length approach, we have also investigated
the contributions of the drag and diffusion
coefficients, resulting from the Born-like scattering : 
$\Lambda_c\pi\rightarrow\Sigma_c\rightarrow\Lambda_c\pi$. 
Using the effective hadronic Lagrangian, \cite{Ko}
\be
{\cal L}_{\Lambda_c\Sigma_c\pi}=\frac{g}{m_\pi}{\overline{\Lambda_c}}
\gamma^5\gamma^\mu{\rm Tr}(\vec{\tau}\cdot\vec{\Sigma}_c\vec{\tau}\cdot\vec{\pi})
+~~{\rm h.c.}~,
\ee
we can calculate the matrix elements for 
$\Lambda_c\pi$ scattering diagrams via $\Sigma_c$.
The Lagrangian is based on the gauged $SU(4)$ flavor symmetry but with empirical
masses. The coupling constant $g=0.37$ is taken from Ref.~\cite{Ko}, where
$SU(4)$ relations are used to fix it.
\begin{figure}
\begin{center}
\includegraphics[scale=0.5]{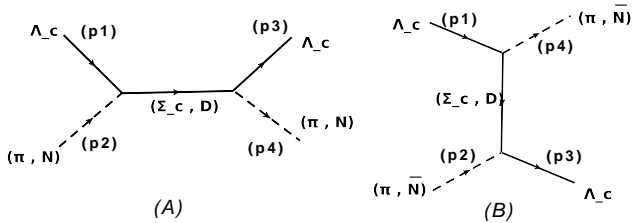}
\caption{Feynman diagrams for the scattering of $\Lambda_c$ with
pion, nucleon and anti-nucleon in the medium.}
\label{fig1}
\end{center}
\end{figure}

The possible $s$ and $u$ channel diagrams 
for $\Lambda_c +\pi\rightarrow \Sigma_c\rightarrow\Lambda_c +\pi$
processes are shown in the panels (A) and (B) of Fig.\ref{fig1}.
The matrix elements for the two channels are respectively
given by,
\be
M^{\Lambda_c\pi}_s=-\left(\frac{2g}{m_\pi}\right)^2\left[ \overline{u}(p_3)\gamma^5\ps_4
\frac{(\ps_1+\ps_2+m_{\Sigma_c})}{(s-m_{\Sigma_c}^2)}
\gamma^5\ps_2 u(p_1)\right]
\ee
and
\be
M^{\Lambda_c\pi}_u=-\left(\frac{2g}{m_\pi}\right)^2\left[ \overline{u}(p_3)\gamma^5\ps_2
\frac{(\ps_1-\ps_4+m_{\Sigma_c})}{(u-m_{\Sigma_c}^2)}
\gamma^5\ps_4 u(p_1)\right]~.
\ee

Similarly from the Lagrangian density~\cite{Ko},
\be
{\cal L}_{\Lambda_cND}=\frac{f}{m_D}{\overline N}
\gamma^5\gamma^\mu\Lambda_c\del_\mu D 
+\del_\mu {\overline D}{\overline{\Lambda_c}}
\gamma^5\gamma^\mu N~,
\ee
one can obtain the matrix elements for 
the processes $\Lambda_cN\rightarrow D\rightarrow\Lambda_cN$
and $\Lambda_c{\overline N}\rightarrow D\rightarrow\Lambda_c{\overline N}$,
(see Fig.\ref{fig1}).
The modulus square of the spin averaged total amplitudes 
${\overline {|{\cal{M}}|^2}}$ for all processes are given in the appendix.
Using those ${\overline {|{\cal{M}}|^2}}$ from effective hadronic model
as well as the corresponding BE and FD distributions for $H=\pi$ and $N$ 
in Eq.(\ref{transport}), 
we can get an alternative estimates of the drag and diffusion coefficients
of $\Lambda_c$ in the hadronic medium. 
We have included form factors in each of the interaction vertices 
to take into account the finite size of the hadrons. For the $u$ and 
$s$-channel diagrams the form factors are taken as~\cite{Ko}
$F_u=\Lambda^2/(\Lambda^2+\vec{q}^2)$  and
$F_s=\Lambda^2/(\Lambda^2+\vec{p_i}^2)$ respectively, where
$\vec{q}$ is the three momentum transfer, $p_i$ is
the initial three momentum of the pions and $\Lambda=1$ GeV.

\section{Results and discussions}
\begin{figure}
\begin{center}
\includegraphics[width=17pc,clip=true]{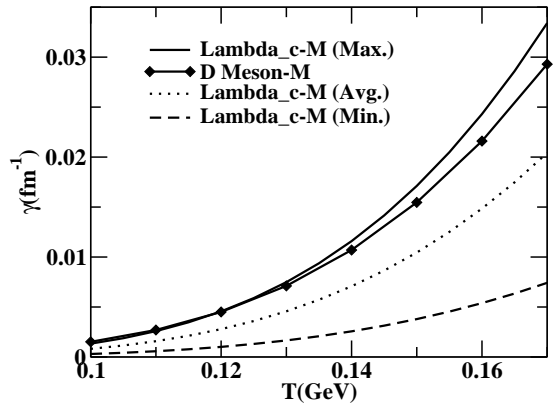}\hspace{2pc}
\caption{Variation of the drag coefficient with temperature for $\Lambda_c$ and 
$D$ meson~\cite{Das3} in a mesonic medium, where their interaction
strengths are governed from their scattering lengths (SL)~\cite{Liu_Zhu}.}
\label{fig2}
\end{center}
\end{figure}

Let us first discuss the results of the drag coefficients obtained 
from the $T$-matrix elements of $\Lambda_cM$ scattering, given in Table ~\ref{table}. 
The variation of the drag coefficient of $\Lambda_c$  with temperature is 
depicted in Fig~\ref{fig2} and compared with the drag coefficient of the 
$D$ mesons~\cite{Das3} while propagating through the same thermal medium consisting of pions, 
kaons and eta. The magnitude of the drag coefficients is quite significant,
indicating a substantial interaction of $\Lambda_c$ 
with the thermal hadrons. 
The maximum and minimum values of the drag coefficient for $\Lambda_c$ 
correspond to the band associated with the $T$ matrix element presented in Table ~\ref{table}. 
The average value of the drag of $\Lambda_c$ is found to be smaller than that of $D$. 

The single electron spectra originating from the decays of $\Lambda_c$ 
and $D$ measured in HIC is sensitive to the following two mechanisms:
(i) the production  of $\Lambda_c$ in HIC is enhanced compared to that in 
pp because of the direct interaction of $c$ with $[ud]$ bound states 
available in the QGP~\cite{lee}, (ii) the $\Lambda_c$ has smaller branching ratio 
to semileptonic decay than $D$. These two mechanisms
lead to a deficiency of electrons at intermediate $p_T$ ($ 2 < p_T$ (GeV) $ < 5$)~\cite{soren}.
If the drag of $\Lambda_c$ is more (less) than $D$ then that will further reduce (enhance) the electrons in this
domain of $p_T$. We find here that the value of the drag of $\Lambda_c$ has a band of uncertainties 
as shown in Fig.~\ref{fig2}, therefore, it is not possible to draw a  conclusion regarding  which way the drag
of $\Lambda_c$ will contribute to the electron spectra originating from the decays of charm mesons and baryons.  
However,  measurements of $D$ meson spectra via hadronic as well as semileptonic
channels in the same collision conditions will help in estimating the electron spectra from 
$\Lambda_c$ and hence its drag coefficients.  

For a hadronic system of life time, $\Delta\tau$ and drag, $\gamma$ 
the momentum suppression is approximately given by, $R_{AA}\sim e^{-\Delta\tau\,\gamma}$~\cite{MinHe}.
Picking up a value of $\gamma$ of $D$ at $T=170$ MeV  from the results displayed
in Fig.~\ref{fig2} we get $R_{AA}\sim 14\%$ for $\Delta\tau=5$ fm/c. The values of $R_{AA}$ for $\Lambda_c$ at the
same temperature and $\Delta\tau$ are about $16\%$ and $4\%$ respectively for maximum and minimum values of $\gamma$  
shown in Fig.\ref{fig2}. Similarly the $\Lambda_c/D$ ratio at the same temperature 
and $\Delta\tau$ is approximately given by, $\Lambda_c/D \sim e^{\Delta\tau\,(\gamma_D-\gamma_{\Lambda_c})}$, 
where $\gamma_D$ and $\gamma_{\Lambda_c}$ are the drag coefficients of the $D$ meson and $\Lambda_c$ respectively.
The $\Lambda_c/D$ ratio can vary upto $12\%$ depending on the minimum to maximum value of 
the drag coefficients of $\Lambda_c$.

\begin{figure}
\begin{center}
\includegraphics[width=17pc,clip=true]{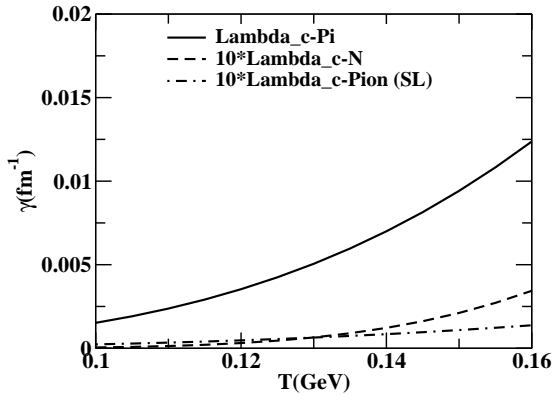}\hspace{2pc}
\caption{Variation of the drag coefficient with temperature for $\Lambda_c$ in a pionic medium,
using the dynamics of Effective Lagrangian (EL) and Scattering Length (SL). The contribution
of drag coefficient for $\Lambda_cN$ scattering in the EL dynamics is
also presented.}
\label{fig3}
\end{center}
\end{figure}
\begin{figure}
\begin{center}
\includegraphics[width=17pc,clip=true]{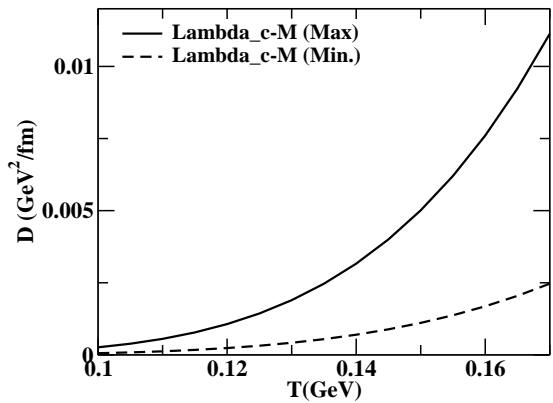}\hspace{2pc}
\caption{Variation of the diffusion coefficient with temperature,
when $\Lambda_c$ interact with all the light pseudo scalar mesons $M$
(in SL dynamics)}
\label{fig4}
\end{center}
\end{figure}

The temperature variation of the drag coefficient  
of $\Lambda_c$ in a pionic medium has been depicted in
Fig.~\ref{fig3}. Here EL corresponds to the matrix 
element obtained from the effective hadronic Lagrangian ~\cite{Ko} and SL 
corresponds to the scattering length or the $T$-matrix element obtained from the  
HB$\chi$PT. 
We  found that the drag of $\Lambda_c$ in pionic medium
for EL is much larger than that for SL. 

The corresponding SL results for momentum diffusion coefficient as
a function of temperature is depicted in Fig.~\ref{fig4}. 
The difference between the maximum and minimum values of both coefficients
get larger at higher temperature as displayed in Figs.~\ref{fig2} and \ref{fig4}. 

\begin{figure}
\begin{center}
\includegraphics[width=17pc,clip=true]{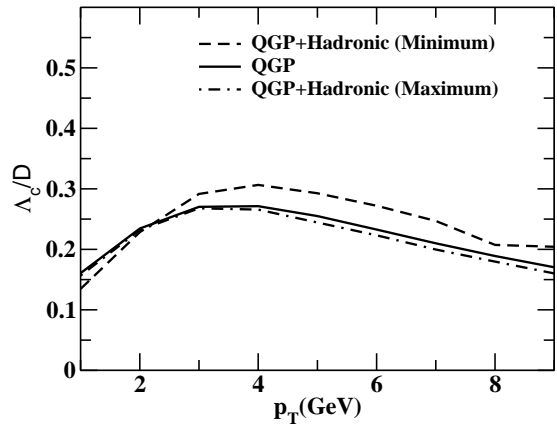}\hspace{2pc}
\caption{Transverse momentum variation of $\Lambda_c$ to $D$ ratio 
has been displayed for maximum and minimum values of the 
drag of $\Lambda_c$ (see text).  
}
\label{fig5}
\end{center}
\end{figure}
The variation of the ratio of $\Lambda_c$ to $D$  is shown
in Fig.~\ref{fig5} as a function of $p_T$.
The Fokker Planck (FP) equation has been used to study the time evolution of 
the $D$ and $\Lambda_c$ in the hadronic 
bath of the equilibrated degrees of freedom. This is given by ~\cite{BS,prdh},  
\be
\frac{\partial f}{\partial t} = 
\frac{\partial}{\partial p_i} \left[ A_i(p)f + 
\frac{\partial}{\partial p_j} \lbrack B_{ij}(p) f\rbrack \right] 
\label{FP}
\label{eq7}
\ee
where $f$ is the momentum distribution of the 
non-equilibrated degrees of freedom , $A_i(p)$ and $B_{ij}(p)$ are related to the drag and diffusion coefficients. 
The interaction between the probe and the thermal bath enter through the 
drag and diffusion coefficients. 
The initial distribution of $D$ meson and $\Lambda_c$ are obtained (at the end of QGP phase) 
by using the fragmentation and coalescence techniques of~\cite{vg} and results of ~\cite{hvh}.
Their ratio (at the end of the QGP phase) has been shown in the solid line of Fig.~\ref{fig5}.
In the present calculation of $\Lambda_c$ spectra, resonances are not taken into account 
at variance with ref.~\cite{lee1}.

The ratio estimated after 
evolving the $D$~\cite{prdh} and $\Lambda_c$ in the hadronic medium
through the Fokker Planck equation is  displayed  in Fig.~\ref{fig5}. 
In Fig.~\ref{fig5} 
QGP refers to the ratio at the end of the QGP phase and ''QGP+Hadronic'' 
refers to the ratio at the end of the Hadronic phase. Maximum and minimum 
of the ratio corresponds to the maximum and minimum values of the drag and diffusion
coefficients of $\Lambda_c$.  The results
indicate that the ratio gets enhanced for  $2\leq p_T\leq 7$ due to 
the  interactions of the $D$ and $\Lambda_c$ while propagating through the
hadronic medium. Such enhancement will have interesting consequences
on the nuclear suppression of the charm quarks in QGP measured through
the single electron spectra originating from the decays of charmed hadrons. 


\section{Summary and discussions}
We have studied the diffusion of  $\Lambda_c$ in a hot hadronic medium.
Using scattering amplitudes, obtained by Liu et al.~\cite{Liu_Zhu}
in the framework of HB$\chi$PT, 
we have evaluated the drag and diffusion coefficients of the $\Lambda_c$ interacting 
with a hadronic background composed of pions,kaons and eta.
We have also calculated the drag 
coefficients of the $\Lambda_c$ interacting  with the pion and nucleon, 
using an  effective 
hadronic Lagrangian. It is found that the coefficients in the pionic medium, 
obtained from the effective hadronic Lagrangian is quite higher than that
obtained from the dynamics of scattering length. However, the coefficients
resulting from the $\Lambda_cN$ scattering obtained within a effective hadronic model 
approach are comparable to the coefficients estimated in the scattering
length approach. The value obtained for 
the $\Lambda_c$ has been compared with the drag coefficient of the $D$ meson calculated 
within the framework of heavy meson chiral perturbation theory (HM$\chi$PT). It is found 
that the value of the drag coefficient of $\Lambda_c$ is generally lower than  that of $D$ mesons.  
This result shows a significant
effect on the $p_T$ dependence of $\Lambda_c/D$ ratio and hence also on $R_{AA}$ of single electrons
originating from the decay of $\Lambda_c$.


\vspace{2mm}
\section*{Acknowledgments}
S.G. thankful to Z.W. Liu and S.L. Zhu
for their useful suggestions via email.
S.G. is supported by FAPESP, Grant n. 2012/16766-0.
S.K.D and V.G acknowledge the support by the ERC StG under the QGPDyn
Grant n. 259684.

\section{Appendix}
The modulus square of the spin averaged total amplitude
for the processes of $\Lambda_c +\pi\rightarrow \Sigma_c\rightarrow\Lambda_c +\pi$
is given by
\bea
{\overline {|M^{\Lambda_c\pi}|^2}}&=&\frac{3}{2}\left(\frac{ 2g}{m_\pi}\right)^4
\left[\frac{A_{ss}}{(s-m_{\Sigma_c}^2)^2}+\frac{A_{uu}}{(u-m_{\Sigma_c}^2)^2}
\right.\nn\\
&&\left.+\frac{2A_{su}}
{(s-m_{\Sigma_c}^2)(u-m_{\Sigma_c}^2)}\right]
\label{M2_Lambda_c}
\eea

where
\bea
A_{ss}&=&[-2m_\pi^2m_{\Lambda_c}(s-m_{\Lambda_c}^2)^2(m_{\Lambda_c}+2m_{\Sigma_c})
\nn\\
&&+m_\pi^4(s+m_{\Lambda_c}^2+2m_{\Lambda_c}m_{\Sigma_c})^2
\nn\\
&&-(s-m_{\Lambda_c}^2)^2(su-m_{\Lambda_c}^4+tm_{\Sigma_c}^2)]~,
\eea
\bea
A_{uu}&=&[-2m_\pi^2m_{\Lambda_c}(u-m_{\Lambda_c}^2)^2(m_{\Lambda_c}+2m_{\Sigma_c})
\nn\\
&&+m_\pi^4(u+m_{\Lambda_c}^2+2m_{\Lambda_c}m_{\Sigma_c})^2
\nn\\
&&-(u-m_{\Lambda_c}^2)^2(su-m_{\Lambda_c}^4+tm_{\Sigma_c}^2)]~,
\eea
and
\bea
A_{su}&=&[-4m_\pi^6m_{\Lambda_c}^2+(4s-2t+4u)m_{\Lambda_c}^6-2m_{\Lambda_c}^8
\nn\\
&&+2t(s^2-t^2-2su+u^2)m_{\Lambda_c}m_{\Sigma_c}+8t^2m_{\Lambda_c}^3m_{\Sigma_c}
\nn\\
&&+m_{\Lambda_c}^4\{-3s^2+5t^2+2s(t-3u)+2tu-3u^2-2tm_{\Sigma_c}^2\}
\nn\\
&&-(s^2-t^2+u^2)(su-tm_{\Sigma_c}^2)+m_{\Lambda_c}^2\{s^3+6t^2m_{\Sigma_c}^2
\nn\\
&&-(t-u)(t+u)^2+s^2(t+3u)-s(t^2+4tu-3u^2)\}
\nn\\
&&+2m_\pi^4\{su+3m_{\Lambda_c}^4-4tm_{\Lambda_c}m_{\Sigma_c}+8m_{\Lambda_c}^3m_{\Sigma_c}
\nn\\
&&-2tm_{\Sigma_c}^2-m_{\Lambda_c}^2(t+4m_{\Sigma_c}^2)\}-2m_\pi^2\{m_{\Lambda_c}m_{\Sigma_c}\{s^2
\nn\\
&&-4t^2+s(t-2u)+tu+u^2\}+14tm_{\Lambda_c}^3m_{\Sigma_c}
\nn\\
&&+8m_{\Lambda_c}^4(t+m_{\Sigma_c}^2)+2t(su-tm_{\Sigma_c}^2)+m_{\Lambda_c}^2\{s^2-2t^2
\nn\\
&&-tu+u^2-s(t+2u)+(-4s+6t-4u)m_{\Sigma_c}^2\}\}]~.
\eea

The modulus square of the spin averaged total amplitude
for the processes of 
$\Lambda_c +N\rightarrow D\rightarrow\Lambda_c +N$
and $\Lambda_c +{\overline N}\rightarrow D\rightarrow\Lambda_c +{\overline N}$
are respectively given by
\bea
{\overline {|M^{\Lambda_cN}|^2}}&=&\frac{1}{2}\left(\frac{f}{m_D}\right)^4
\frac{1}{(s-m_D^2)^2}
\nn\\
&&{\rm Tr}[(\ps_3+m_{\Lambda_c})(\ps_1+\ps_2)
(\ps_4-m_N)(\ps_1+\ps_2)]
\nn\\
&&{\rm Tr}[(\ps_2-m_N)(\ps_1+\ps_2)
(\ps_1+m_{\Lambda_c})(\ps_1+\ps_2)]
\nn\\
&=&\frac{2(f/m_D)^4}{(s-m_D^2)^2}(m_{\Lambda_c}-m_N)^2
\{s-(m_{\Lambda_c}+m_N)^2\}
\nn\\
&&[3(m_{\Lambda_c}^4+m_N^4)+10m_{\Lambda_c}^2m_N^2+(t+u)^2-s^2
\nn\\
&&-4(t+u)(m_{\Lambda_c}^2+m_N^2)+s(m_{\Lambda_c}-m_N)^2]
\nn\\
\eea
and
\bea
{\overline {|M^{\Lambda_c{\overline N}}|^2}}&=&\frac{1}{2}\left(\frac{f}{m_D}\right)^4
\frac{1}{(u-m_D^2)^2}
\nn\\
&&{\rm Tr}[(\ps_3+m_{\Lambda_c})(\ps_1-\ps_4)
(\ps_2+m_N)(\ps_1-\ps_4)]
\nn\\
&&{\rm Tr}[(\ps_1+m_{\Lambda_c})(\ps_1-\ps_4)
(\ps_4+m_N)(\ps_1-\ps_4)]
\nn\\
&=&\frac{2(f/m_D)^4}{(u-m_D^2)^2}(m_{\Lambda_c}-m_N)^2
\{u-(m_{\Lambda_c}+m_N)^2\}
\nn\\
&&[3(m_{\Lambda_c}^4+m_N^4)+10m_{\Lambda_c}^2m_N^2+(t+s)^2-u^2
\nn\\
&&-4(t+s)(m_{\Lambda_c}^2+m_N^2)+u(m_{\Lambda_c}-m_N)^2]~.
\nn\\
\eea

\end{document}